\documentstyle[12pt]{article}
\begin{document}
\baselineskip=15.5pt
\begin{titlepage}

\begin{flushright}
IC/2002/4\\
hep-th/0201268
\end{flushright}
\vspace{10 mm}

\begin{center}
{\Large  A Note on the Cardy-Verlinde Formula}

\vspace{5mm}

\end{center}

\vspace{5 mm}

\begin{center}
{\large Donam Youm\footnote{E-mail: youmd@ictp.trieste.it}}

\vspace{3mm}

ICTP, Strada Costiera 11, 34014 Trieste, Italy

\end{center}

\vspace{1cm}

\begin{center}
{\large Abstract}
\end{center}

\noindent

We generalize the results of hep-th/0008140 to the case of the 
$(n+1)$-dimensional closed FRW universe satisfying a general equation 
of state of the form $p=w\rho$.  We find that the entropy of the universe 
can no longer be expressed in a form similar to the Cardy formula, when 
$w\neq 1/n$.  As a result, in general the entropy formula does not coincide 
with the Friedmann equation when the conjectured bound on the Casimir energy 
is saturated.  Furthermore, the conjectured bound on the Casimir energy 
generally does not lead to the Hubble and the Bekenstein entropy bounds.  

\vspace{1cm}
\begin{flushleft}
January, 2002
\end{flushleft}
\end{titlepage}
\newpage

Verlinde \cite{ver1} made an interesting proposal that the Cardy formula 
\cite{car} for the two-dimensional CFT can be generalized to arbitrary 
spacetime dimensions.  Verlinde further proposed that a closed universe 
has subextensive (Casimir) contribution to its energy and entropy with the 
Casimir energy conjectured to be bounded from above by the Bekenstein-Hawking 
energy. Within the context of the radiation dominated universe, such bound on 
the Casimir energy is shown to lead to the Hubble and the Bekenstein entropy 
bounds respectively for the strongly and the weakly self-gravitating 
universes.  The generalized entropy formula, called the Cardy-Verlinde 
formula, is further shown to coincide with the Friedmann equation at the 
moment when the conjectured bound on the Casimir energy is saturated.  
These results were later generalized 
\cite{was,kps1,cai,bm,bim,kps2,youm,cai2,noo,caz,odi1,odi2,youm2,waa,nooq,dan,ogu,pes,rgc,med1,med2,youm3,hal,myu,amed,rcai,jin} 
to brane universes in the bulks of various (A)dS black holes.  

It is the purpose of this note to examine whether the above mentioned results 
of Ref. \cite{ver1} are generic properties of the $(n+1)$-dimensional closed 
Friedmann-Robertson-Walker (FRW) universe satisfying a general equation of 
state of the form $p=w\rho$.  We find that the entropy of the FRW universe 
can no longer be expressed in a form similar to the Cardy formula with 
square root, when $w\neq 1/n$.  As a result, the entropy expression does 
not coincide with the Friedmann equation at the moment when the conjectured 
bound on the Casimir energy is saturated and the conjectured bound on the 
Casimir energy does not lead to the Hubble and the Bekenstein entropy bounds.

The general metric ansatz for an $(n+1)$-dimensional universe satisfying 
the principles of homogeneity and isotropy in the $n$-dimensional space 
has the following Robertson-Walker form:
\begin{equation}
g_{\mu\nu}dx^{\mu}dx^{\nu}=-d\tau^2+a^2(\tau)\gamma_{ij}dx^idx^j,
\label{rwmet}
\end{equation}
where $a$ is the cosmic scale factor and $\gamma_{ij}$ is given by
\begin{equation}
\gamma_{ij}dx^idx^j=\left(1+\textstyle{k\over 4}\delta_{mn}x^mx^n\right)^{-2}
\delta_{ij}dx^idx^j={{dr^2}\over{1-kr^2}}+r^2d\Omega^2_{n-1}.
\label{maxmet}
\end{equation}
Here, $k=-1,0,1$, respectively for the $n$-dimensional space with the 
negative, zero and positive spatial curvature.  In this paper, we mostly 
concentrate on the closed universe case ($k=1$).  Generalization to the 
other cases is straightforward.  

The Friedmann equations, describing the expansion of the $(n+1)$-dimensional 
universe described by the Robertson-Walker metric (\ref{rwmet}), are given by
\begin{equation}
H^2={{16\pi G}\over{n(n-1)}}\rho-{k\over a^2},
\label{frd1}
\end{equation}
\begin{equation}
\dot{H}=-{{8\pi G}\over{n-1}}(\rho+p)+{k\over a^2},
\label{frd2}
\end{equation}
where $H\equiv \dot{a}/a$ is the Hubble parameter, and $\rho$ and $p$ are 
the energy density and the pressure of the universe.  
From the Friedmann equations, we obtain the following energy conservation 
equation:
\begin{equation}
\dot{\rho}+n(\rho+p){\dot{a}\over a}=0,
\label{encons}
\end{equation}
where the overdot denotes derivative w.r.t. $\tau$.  

For the perfect fluid matter of the universe satisfying an equation of 
state
\begin{equation}
p=w\rho,
\label{eqst}
\end{equation}
by solving the energy conservation equation (\ref{encons}) we obtain
\begin{equation}
\rho\propto a^{-n(1+w)}.
\label{rhoa}
\end{equation}
We denote the constant of proportionality in this equation as $\rho_0$, 
namely, $\rho=\rho_0a^{-n(1+w)}$.  

We assume that the FRW universe satisfies the first law of thermodynamics.  
When applied to a comoving volume element of unit coordinate volume and 
physical volume $v=a^n$, the first law of thermodynamics takes the form:
\begin{equation}
Tds=d(\rho a^n)+pda^n,
\label{2ndlaw}
\end{equation}
where $T$ is the temperature of the universe and $s$ is the entropy density 
of the universe per comoving volume.  Making use of Eqs. (\ref{encons}) 
and (\ref{2ndlaw}), we see that entropy of the FRW universe per comoving 
volume stays constant in time, i.e., $ds/d\tau=0$.  Therefore, {\it the FRW 
universe satisfying the first law of thermodynamics expands adiabatically}: 
$dS=0$, where $S=s\int d^nx\sqrt{\gamma}$ is the total entropy inside the 
total volume $V=a^n\int d^nx\sqrt{\gamma}$ of the universe. 
By substituting the following relation, obtained from the integrability 
condition $\partial^2s/(\partial T\partial v)=\partial^2s/(\partial v
\partial T)$:
\begin{equation}
{{dp}\over{dT}}={{p+\rho}\over T},
\label{thmrel}
\end{equation}
into Eq. (\ref{2ndlaw}), we obtain the following expression for entropy of 
the FRW universe per comoving volume:
\begin{equation}
s={a^n\over T}(p+\rho)+s_0,
\label{entuniv}
\end{equation}
where $s_0$ is an integration constant.
For a universe satisfying an equation of state of the form $p=w\rho$, 
the temperature of the universe is therefore given by
\begin{equation}
T={{(1+w)\rho_0}\over{s-s_0}}a^{-nw}.
\label{temppwr}
\end{equation}

The total energy $E=\rho V$ can be written as a sum of the purely extensive 
part $E_E$ and the subextensive part $E_C$, called the Casimir energy, in 
the following way:
\begin{equation}
E(S,V)=E_E(S,V)+{1\over 2}E_C(S,V).
\label{toten}
\end{equation}
Under the transformations $S\to\lambda S$ and $V\to\lambda V$ with a constant 
$\lambda$, the extensive and the subextensive parts of the total energy 
respectively scale as
\begin{eqnarray}
E_E(\lambda S,\lambda V)&=&\lambda E_E(S,V),
\cr
E_C(\lambda S,\lambda V)&=&\lambda^{1-2/n}E_C(S,V).
\label{casenscl}
\end{eqnarray}
Therefore, we have $E(\lambda S,\lambda V)=\lambda E_E(S,V)+{1\over 2}
\lambda^{1-2/n}E_C(S,V)$.  Taking derivative of this relation w.r.t. $\lambda$ 
and letting $\lambda=1$, we obtain
\begin{equation}
S\left({{\partial E}\over{\partial S}}\right)_V+V\left({{\partial E}\over
{\partial V}}\right)_S=E_E+\left({1\over 2}-{1\over n}\right)E_C.
\label{intrel}
\end{equation}
Since we assume that the universe satisfies the first law of thermodynamics 
$dE=TdS-pdV$, we have the thermodynamics relations $\left({{\partial E}\over
{\partial V}}\right)_S=-p$ and $\left({{\partial E}\over{\partial S}}
\right)_V=T$.  Making use of these thermodynamic relations and Eq. 
(\ref{toten}), we can put Eq. (\ref{intrel}) into the following definition 
for the Casimir energy as the violation of the Euler identity:
\begin{equation}
E_C=n(E+pV-TS).
\label{casendef}
\end{equation}
By comparing Eq. (\ref{entuniv}) with Eq. (\ref{casendef}), we see that  
the Casimir energy of the closed FRW universe has the form
\begin{equation}
E_C=-nTs_0\int d^nx\sqrt{\gamma}.
\label{univcasen}
\end{equation}
For the universe satisfying the equation of state $p=w\rho$, the Casimir 
energy therefore behaves with the cosmic scale factor like $E_C\sim a^{-nw}$.  
Since $E\sim\rho a^n\sim a^{-nw}$, the extensive part $E_E=E-{1\over 2}E_C$ 
of energy also goes like $E_E\sim a^{-nw}$.  Unlike the case of the radiation 
dominated FRW universe, the Casimir entropy $S_C={{2\pi}\over n}E_Ca\sim 
a^{1-nw}$ does not remain constant during the cosmological evolution, when 
$w\neq 1/n$.  Making use of the above properties of $E_C$ and $E_E$, in the 
following we obtain the expression for the total entropy $S$ of the universe.

We begin by reviewing the case considered in Ref. \cite{ver1}, namely the 
radiation dominated universe, for which the universe has conformal invariance. 
If we assume the conformal invariance, the products $E_Ea$ and $E_Ca$ are 
independent of the volume $V\sim a^n$ and are functions of the entropy $S$, 
only, since $E_E\sim a^{-1}$ and $E_C\sim a^{-1}$ for $w=1/n$.  
Making use of this fact and the rescaling behaviors (\ref{casenscl}) of 
$E_E$ and $E_C$, we infer the following expressions \cite{ver1}:
\begin{equation}
E_E={\alpha\over{4\pi a}}S^{1+1/n},\ \ \ \ \ \ \ 
E_C={\beta\over{2\pi a}}S^{1-1/n},
\label{rades}
\end{equation}
where $\alpha$ and $\beta$ are arbitrary constants.  Making use of Eqs. 
(\ref{toten}) and (\ref{rades}), we obtain the following expression 
\cite{ver1} for entropy resembling the Cardy formula
\footnote{In general for a universe with the curvature parameter $k$, 
Eq. (\ref{toten}) should be replaced by $E=E_E+{k\over 2}E_C$.  With this 
general expression for $E$, we obtain $S={{2\pi a}\over\sqrt{\alpha\beta}}
\sqrt{E_C(2E-kE_C)}$ and Eq. (\ref{casendef}) is replaced by $kE_C=n(E+pV-
TS)$, which are exactly the results obtained in Ref. \cite{youm}.} 
\cite{car}:
\begin{equation}
S={{2\pi a}\over\sqrt{\alpha\beta}}\sqrt{E_C(2E-E_C)}.
\label{cfent}
\end{equation} 
The undetermined normalization factor $\sqrt{\alpha\beta}$ is fixed 
\cite{ver1} to be $n$ for CFT with an AdS dual, making use of the fact that 
thermodynamic quantities of the CFT at high temperature can be identified 
with those of the bulk AdS black hole \cite{wit}.  

We now consider more general case.  For the universe satisfying an equation 
of state (\ref{eqst}) with a general $w$, the Cardy-Verlinde formula 
(\ref{cfent}) no longer holds.  This can be easily seen by the fact that 
with $E\sim a^{-nw}$ and $E_C\sim a^{-nw}$ the entropy expression in Eq. 
(\ref{cfent}) goes like $S\sim a^{1-nw}$, which is contradictory with the 
fact that $S=s\int d^nx\sqrt{\gamma}$ is constant in time for the FRW 
universe satisfying the first law of thermodynamics.  
For an arbitrary value of $w$, it is rather the products $E_Ca^{nw}$ and 
$E_Ea^{nw}$ that are independent of $V$ and should be functions of $S$, only.  
Making use of the general scaling behaviors (\ref{casenscl}), we therefore 
obtain the following expressions for the extensive and the subextensive 
parts of the total energy of the FRW universe satisfying $p=w\rho$:
\begin{equation}
E_E={\alpha\over{4\pi a^{nw}}}S^{w+1}, \ \ \ \ \ \ \ 
E_C={\beta\over{2\pi a^{nw}}}S^{w+1-2/n},
\label{genens}
\end{equation}
where $\alpha$ and $\beta$ are again arbitrary constants.  
From these expressions for $E_E$ and $E_C$, we obtain the following 
expression for the entropy of the universe:
\begin{equation}
S=\left[{{2\pi a^{nw}}\over\sqrt{\alpha\beta}}\sqrt{E_C(2E-E_C)}
\right]^{n\over{(w+1)n-1}}.
\label{genent}
\end{equation}
We therefore see that the expression for entropy of the universe resembling 
the Cardy formula (with square root) is special only for the radiation 
dominated universe ($w=1/n$).  This new entropy formula can be expressed 
as the following entropy relation:
\begin{equation}
S^{2(w-1/n+1)}={n^2\over{\alpha\beta}}a^{2(nw-1)}(2S_BS_C-S^2_C),
\label{entrel}
\end{equation}
where $S_B\equiv{{2\pi a}\over n}E$ is the Bekenstein entropy.  So, 
clean relation among various entropies is possible only for the 
$w=1/n$ case.   Note, the above entropy expression holds only for the 
case in which the universe expands adiabatically.  While the entropy 
increases ($dS>0$) as the universe expands, the entropy will not be 
expressed solely in terms of $E$ and $E_C$ as above.  

The Friedmann equations (\ref{frd1}) and (\ref{frd2}) can be expressed in the 
following forms resembling thermodynamic formulas of CFT, regardless of the 
form of equation of state satisfied by the universe:
\begin{equation}
S_H={{2\pi}\over n}a\sqrt{E_{BH}(2E-kE_{BH})},
\label{thrfm1}
\end{equation}
\begin{equation}
kE_{BH}=n(E+pV-T_HS_H),
\label{thrfm2}
\end{equation}
in terms of the Hubble entropy $S_H$ and the Bekenstein-Hawking energy 
$E_{BH}$, where
\begin{equation}
S_H\equiv (n-1){{HV}\over{4G}},\ \ \ \ \ 
E_{BH}\equiv n(n-1){V\over{8\pi Ga^2}},\ \ \ \ \ 
T_H\equiv -{\dot{H}\over{2\pi H}}.
\label{defs}
\end{equation}
Eq. (\ref{thrfm1}) is referred to as the cosmological Cardy formula, due to 
its resemblance to the Cardy formula for the two-dimensional CFT.  Note, 
Eq. (\ref{thrfm2}) resembles the Smarr's formula for a thermodynamics system 
having the Casimir contribution.  The first Friedmann equation (\ref{frd1}) 
can be expressed also as the following relation among the Bekenstein entropy 
$S_B\equiv{{2\pi a}\over n}E$, the Bekenstein-Hawking entropy $S_{BH}\equiv
(n-1){V\over{4Ga}}$ and the Hubble entropy $S_H$:
\begin{equation}
S^2_H=2S_BS_{BH}-kS^2_{BH}.
\label{frd1entrel}
\end{equation}
This can be expressed as the following quadratic relation, when $k=1$:
\begin{equation}
S^2_H+(S_B-S_{BH})^2=S^2_B.
\label{entquad}
\end{equation}
The Bekenstein entropy goes like $S_B\sim Ea\sim a^{1-nw}$ and therefore 
remains constant during the cosmological evolution only for the $w=1/n$ case.  

We assume that the cosmological bound on the Casimir energy $E_C$, proposed 
by Verlinde \cite{ver1}, continues to hold even when the universe is not 
radiation dominated:
\begin{equation}
E_C\leq E_{BH}.
\label{cshlbnd}
\end{equation}
For a universe satisfying the equation of state $p=w\rho$, we have 
seen that $E_C$ behaves like $a^{-nw}$, whereas $E_{BH}$ always goes like  
$a^{n-2}$.  This implies that the cosmological bound (\ref{cshlbnd}) 
is satisfied, provided the universe is larger [smaller] than a certain 
critical size (for which the bound is saturated) when $w>-(n-2)/n$ 
[$w<-(n-2)/n$].  In particular, the radiation dominated universe 
($w=1/n$) and the matter dominated universe ($w=0$) correspond to the 
former case, and the vacuum dominated universe ($w=-1$) corresponds to 
the latter case.  For the $w=-(n-2)/n$ case, the bound can never be 
satisfied depending on the values of the constant coefficients of $E_C$ and 
$E_{BH}$.  

It is shown \cite{ver1,ver2} that the Friedmann equations for the radiation 
dominated universe ($w=1/n$) coincide with thermodynamic formulae of the 
dual CFT at the moment when the cosmological bound (\ref{cshlbnd}) on $E_C$ 
is saturated.  However, for $w\neq 1/n$, the modified Cardy-Verlinde formula 
(\ref{genent}) and the cosmological Cardy formula (\ref{thrfm1}) with $k=1$ 
do not coincide when $E_C=E_{BH}$.  Therefore, the matching of the first 
Friedmann equation and the Cardy-Verlinde formula is accidental for the 
radiation dominated universe, for which entropy of the universe is expressed 
in terms of square root just like the Cardy formula.  Next, we consider the 
second Friedmann equation.  Although $S$ and $T$ do not respectively reduce to 
$S_H$ and $T_H$ when $E_C=E_{BH}$, it can be shown by using Eqs. (\ref{frd2}) 
and (\ref{casendef}) that the product $T_HS_H$ reduces to $TS$ when $E_C=
E_{BH}$, for the $k=1$ case
\footnote{This fact can be shown to hold even for the $k\neq 1$ case by using 
the generalized expression given in the previous footnote.}.  
Therefore, regardless of the form of the equation of state, the cosmological 
Smarr's formula (\ref{thrfm2}) reduces to the Smarr's formula (\ref{casendef}) 
for a thermodynamic system with the Casimir energy $E_C$ when the bound 
(\ref{cshlbnd}) on $E_C$ is saturated.  

We now examine whether the cosmological bound (\ref{cshlbnd}) on $E_C$ 
continues to imply the Hubble entropy bound and the Bekenstein bound even for 
the $w\neq 1/n$ case.  The criteria for a weakly and a strongly 
self-gravitating closed universes are respectively
\begin{eqnarray}
E\leq E_{BH} \ \ \ \ \ \ \ &{\rm for}&\ \ \ \ \ \ \ 
Ha\leq 1
\cr
E\geq E_{BH} \ \ \ \ \ \ \ &{\rm for}&\ \ \ \ \ \ \ 
Ha\geq 1.
\label{wsucrt}
\end{eqnarray}
First, we consider the strongly self-gravitating case.  From Eqs. 
(\ref{cshlbnd}) and (\ref{wsucrt}) we see that $E_C\leq E_{BH}\leq E$.  
Furthermore, we see from Eq. (\ref{genent}) that $S$ is a monotonically 
increasing [decreasing] function of $E_C$ in the interval $E_C\leq E$ 
when $w>-(n-1)/n$ [$w<-(n-1)/n$]
\footnote{When $w=-(n-1)/n$, $S$ is insensitive to variation of $E_C$  
and, therefore, the conjectured bound (\ref{cshlbnd}) on $E_C$ does not 
lead to cosmological entropy bounds.}.  
Therefore, $S$ takes the maximum [minimum] value when $E_C=E$ and 
therefore $E_C=E_{BH}$, for the $w>-(n-1)/n$ [$w<-(n-1)/n$] case.  The 
extremum value of $S$ (when $E_C=E_{BH}$) is given by $S=\left[{n\over
\sqrt{\alpha\beta}}a^{nw-1}S_H\right]^{n\over{(w+1)n-1}}$, as can be seen 
from Eq. (\ref{thrfm1}) with $k=1$ and Eq. (\ref{genent}).  So, the 
cosmological entropy bound for the strongly self-gravitating closed FRW 
universe, resulting from Eq. (\ref{cshlbnd}), is
\begin{eqnarray}
S\leq S_0\left[a^{nw-1}S_H\right]^{n\over{(w+1)n-1}}
\ \ \ \ \ \ \ \ &{\rm for}&\ \ \ \ \ \ \ \ w>-(n-1)/n
\cr
S\geq S_0\left[a^{nw-1}S_H\right]^{n\over{(w+1)n-1}}
\ \ \ \ \ \ \ \ &{\rm for}&\ \ \ \ \ \ \ \ w<-(n-1)/n,
\label{strentbnd}
\end{eqnarray}
where we have absorbed all the constant factors into an undetermined constant 
$S_0$.  
Second, we consider the weakly self-gravitating case.  Since $E\leq E_{BH}$ 
for such case, the maximum [minimum] for $S$ is reached earlier, namely when 
$E_C=E$, for the $w>-(n-1)/n$ [$w<-(n-1)/n$] case.  The extremum (the value 
of $S$ when $E_C=E$) is $S=\left[{{2\pi a^{nw}}\over\sqrt{\alpha\beta}}E
\right]^{n\over{(w+1)n-1}}=\left[{n\over\sqrt{\alpha\beta}}a^{nw-1}
S_B\right]^{n\over{(w+1)n-1}}$.  The cosmological entropy bound for the 
weakly self-gravitating closed FRW universe, resulting from the bound on 
$E_C$, is therefore
\begin{eqnarray}
S\leq S_0\left[a^{nw-1}S_B\right]^{n\over{(w+1)n-1}}
\ \ \ \ \ \ \ \ &{\rm for}&\ \ \ \ \ \ \ \ w>-(n-1)/n
\cr
S\geq S_0\left[a^{nw-1}S_B\right]^{n\over{(w+1)n-1}}
\ \ \ \ \ \ \ \ &{\rm for}&\ \ \ \ \ \ \ \ w<-(n-1)/n.
\label{wkentbnd}
\end{eqnarray}
We have therefore seen that the conjectured bound (\ref{cshlbnd}) on $E_C$ 
implies the Hubble and the Bekenstein bounds only for the radiation 
dominated universe case.  It appears from the entropy expression 
(\ref{genent}) that the Hubble and the Bekenstein entropy bounds are not 
likely to be reproduced when $w\neq 1/n$, even with modified bound on $E_C$.


\begin{thebibliography} {99}
\small
\parskip=0pt plus 2pt

\bibitem{ver1} E. Verlinde, ``On the holographic principle in a radiation 
dominated universe,'' hep-th/0008140.
%%CITATION = HEP-TH 0008140;%%

\bibitem{car} J.L. Cardy, ``Operator content of two-dimensional conformally 
invariant Theories,'' Nucl. Phys. {\bf B270} (1986) 186.
%%CITATION = NUPHA,B270,186;%%

\bibitem{was} B. Wang, E. Abdalla and R. Su, ``Relating Friedmann equation 
to Cardy formula in universes with cosmological constant,'' Phys. Lett. 
{\bf B503} (2001) 394, hep-th/0101073.
%%CITATION = HEP-TH 0101073;%%

\bibitem{kps1} D. Klemm, A.C. Petkou and G. Siopsis, ``Entropy bounds, 
monotonicity properties and scaling in CFTs,'' Nucl. Phys. {\bf B601} 
(2001) 380, hep-th/0101076.
%%CITATION = HEP-TH 0101076;%%

\bibitem{cai} R.G. Cai, ``The Cardy-Verlinde formula and AdS black holes,''
Phys. Rev. {\bf D63} (2001) 124018, hep-th/0102113.
%%CITATION = HEP-TH 0102113;%%

\bibitem{bm} A.K. Biswas and S. Mukherji, ``Holography and stiff-matter on 
the brane,'' JHEP {\bf 0103} (2001) 046, hep-th/0102138.
%%CITATION = HEP-TH 0102138;%%

\bibitem{bim} D. Birmingham and S. Mokhtari, ``The Cardy-Verlinde formula and 
Taub-Bolt-AdS spacetimes,'' Phys. Lett. {\bf B508} (2001) 365, hep-th/0103108.
%%CITATION = HEP-TH 0103108;%%

\bibitem{kps2} D. Klemm, A.C. Petkou, G. Siopsis and D. Zanon, 
``Universality and a generalized C-function in CFTs with AdS duals,'' 
Nucl. Phys. {\bf B620} (2002) 519, hep-th/0104141.
%%CITATION = HEP-TH 0104141;%%

\bibitem{youm} D. Youm, ``The Cardy-Verlinde formula and topological 
AdS-Schwarzschild black  holes,'' Phys. Lett. {\bf B515} (2001) 170, 
hep-th/0105093.
%%CITATION = HEP-TH 0105093;%%

\bibitem{cai2} R. Cai, Y.S. Myung, and N. Ohta, ``Bekenstein bound, 
holography and brane cosmology in charged black hole background,'' 
Class. Quant. Grav. {\bf 18} (2001) 5429, hep-th/0105070.
%%CITATION = HEP-TH 0105070;%%

\bibitem{noo} S. Nojiri, S.D. Odintsov and S. Ogushi, ``Holographic entropy 
and brane FRW-dynamics from AdS black hole in d5 higher derivative gravity,'' 
Int. J. Mod. Phys. {\bf A16} (2001) 5085, hep-th/0105117.
%%CITATION = HEP-TH 0105117;%%

\bibitem{caz} R.G. Cai and Y.Z. Zhang, ``Holography and brane cosmology in 
domain wall backgrounds,'' Phys. Rev. {\bf D64} (2001) 104015, hep-th/0105214.
%%CITATION = HEP-TH 0105214;%%

\bibitem{odi1} S. Nojiri and S.D. Odintsov, ``AdS/CFT correspondence, 
conformal anomaly and quantum corrected entropy  bounds,'' Int. J. Mod. 
Phys. {\bf A16} (2001) 3273, hep-th/0011115.
%%CITATION = HEP-TH 0011115;%%

\bibitem{odi2} S. Nojiri and S.D. Odintsov, ``AdS/CFT and quantum-corrected 
brane entropy,'' Class. Quant. Grav. {\bf 18} (2001) 5227, hep-th/0103078.
%%CITATION = HEP-TH 0103078;%%

\bibitem{youm2} D. Youm, ``The Cardy-Verlinde formula and charged topological 
AdS black holes,'' Mod. Phys. Lett. {\bf A16} (2001) 1327, hep-th/0105249.
%%CITATION = HEP-TH 0105249;%%

\bibitem{waa} B. Wang, E. Abdalla and R.K. Su, ``Friedmann equation and 
Cardy formula correspondence in brane universes,'' hep-th/0106086.
%%CITATION = HEP-TH 0106086;%%

\bibitem{nooq} S. Nojiri, O. Obregon, S.D. Odintsov, H. Quevedo and M.P. Ryan, 
``Quantum bounds for gravitational de Sitter entropy and the  Cardy-Verlinde 
formula,'' Mod. Phys. Lett. {\bf A16} (2001) 1181, hep-th/0105052.
%%CITATION = HEP-TH 0105052;%%

\bibitem{dan} U.H. Danielsson, ``A black hole hologram in de Sitter space,'' 
hep-th/0110265.
%%CITATION = HEP-TH 0110265;%%

\bibitem{ogu} S. Ogushi, ``Holographic entropy on the brane in de Sitter 
Schwarzschild space,'' hep-th/0111008.
%%CITATION = HEP-TH 0111008;%%

\bibitem{pes} A.C. Petkou and G. Siopsis, ``dS/CFT correspondence on a 
brane,'' hep-th/0111085.
%%CITATION = HEP-TH 0111085;%%

\bibitem{rgc} R.G. Cai, ``Cardy-Verlinde formula and asymptotically de 
Sitter spaces,'' Phys. Lett. {\bf B525} (2002) 331, hep-th/0111093.
%%CITATION = HEP-TH 0111093;%%

\bibitem{med1} A.J.M. Medved, ``CFT on the brane with a Reissner-Nordstrom-de 
Sitter twist,'' hep-th/0111182.
%%CITATION = HEP-TH 0111182;%%

\bibitem{med2} A.J.M. Medved, ``dS/CFT duality with a topological twist,'' 
hep-th/0111238.
%%CITATION = HEP-TH 0111238;%%

\bibitem{youm3} D. Youm, ``The Cardy-Verlinde formula and asymptotically 
de Sitter brane universe,'' hep-th/0111276.
%%CITATION = HEP-TH 0111276;%%

\bibitem{hal} E. Halyo, ``On the Cardy-Verlinde formula and the de 
Sitter/CFT correspondence,'' hep-th/0112093.
%%CITATION = HEP-TH 0112093;%%

\bibitem{myu} Y.S. Myung, ``Dynamic dS/CFT correspondence using the brane 
cosmology,'' hep-th/0112140.
%%CITATION = HEP-TH 0112140;%%

\bibitem{amed} A.J. Medved, ``A holographic interpretation of asymptotically 
de Sitter spacetimes,'' hep-th/0112226.
%%CITATION = HEP-TH 0112226;%%

\bibitem{rcai} R.G. Cai, ``Cardy-Verlinde formula and thermodynamics of 
black holes in de Sitter  spaces,'' hep-th/0112253.
%%CITATION = HEP-TH 0112253;%%

\bibitem{jin} J. Jing, ``Cardy-Verlinde Formula and Kerr-Newman-AdS$_4$ 
and two rotation parameters Kerr-AdS$_5$ black holes,'' hep-th/0201247.
%%CITATION = HEP-TH 0201247;%%

\bibitem{wit} E. Witten, ``Anti-de Sitter space, thermal phase transition,
and confinement in  gauge theories,'' Adv. Theor. Math. Phys. {\bf 2} (1998)
505, hep-th/9803131.
%%CITATION = HEP-TH 9803131;%%

\bibitem{ver2} I. Savonije and E. Verlinde, ``CFT and entropy on the brane,''
Phys. Lett. {\bf B507} (2001) 305, hep-th/0102042.
%%CITATION = HEP-TH 0102042;%%

\end{thebibliography}
\end{document}